\documentstyle[12pt]{article}
\def\ie{{\em i.e.}}
\def\eg{{\em e.g.}}

\def\beq{\begin{equation}}
\def\eeq{\end{equation}}

\catcode`\@=11 

\def\lsim{\mathrel{\mathpalette\@versim<}}
\def\gsim{\mathrel{\mathpalette\@versim>}}
\def\@versim#1#2{\vcenter{\offinterlineskip
    \ialign{$\m@th#1\hfil##\hfil$\crcr#2\crcr\sim\crcr } }}

\def\JL{J. L. Lopez}
\def\DVN{D. V. Nanopoulos}
\def\AZ{A. Zichichi}

\def\t1{{\tilde 1}}

\def\mpT{p_T\hskip-12pt/\hskip5pt}

\def\GeV{\,{\rm GeV}}

\def\to{\rightarrow}
\def\pb{\,{\rm pb}}
\def\ipb{\,{\rm pb}^{-1}}
\def\ifb{\,{\rm fb}^{-1}}

\def\NPB#1#2#3{Nucl. Phys. B {\bf#1} (19#2) #3}
\def\PLB#1#2#3{Phys. Lett. B {\bf#1} (19#2) #3}

\def\PRD#1#2#3{Phys. Rev. D {\bf#1} (19#2) #3}

\def\MODA#1#2#3{Mod. Phys. Lett. A {\bf#1} (19#2) #3}

\def\TAMU#1{Texas A \& M University preprint CTP-TAMU-#1}

\begin{document}
\begin{flushright}
\baselineskip=12pt
{CERN-TH.7535/94}\\
{CTP-TAMU-66/94}\\
{ACT-23/94}\\
{hep-ph/9412346}\\
\end{flushright}

\begin{center}
{\Huge\bf Supersymmetry dileptons and trileptons at the Tevatron\\}
\vglue 1cm
{JORGE L. LOPEZ$^{1,2}$, D. V. NANOPOULOS$^{1,2,3}$, XU WANG$^{1,2}$,
and A.~ZICHICHI$^{4}$\\}
\vglue 0.4cm
{\em $^{1}$Center for Theoretical Physics, Department of Physics, Texas A\&M
University\\}
{\em College Station, TX 77843--4242, USA\\}
{\em $^{2}$Astroparticle Physics Group, Houston Advanced Research Center
(HARC)\\}
{\em The Mitchell Campus, The Woodlands, TX 77381, USA\\}
{\em $^{3}$CERN Theory Division, 1211 Geneva 23, Switzerland\\}
{\em $^{4}$CERN, 1211 Geneva 23, Switzerland\\}
\baselineskip=12pt
\end{center}

\vglue 0.5cm
\begin{abstract}
We consider the production of supersymmetry neutralinos and charginos in
$p\bar p$ collisions at the Tevatron, and their subsequent decay via
hadronically quiet dileptons and trileptons. We perform our computations in the
context of a variety of supergravity models, including generic four-parameter
supergravity models, the minimal $SU(5)$ supergravity model, and $SU(5)\times
U(1)$ supergravity with string inspired two- and one-parameter moduli and
dilaton scenarios. Our results are contrasted with estimated experimental
sensitivities for dileptons and trileptons for integrated luminosities of
$100\ipb$ and $1\ifb$, which should be available in the short and long terms.
We find that the dilepton mode is a needed complement to the trilepton signal
when the latter is suppressed by small neutralino leptonic branching ratios.
The estimated reaches in chargino masses can be as large as $100\,(150)\GeV$
for $100\ipb\,(1\ifb)$. We also discuss the task left for LEPII once the
Tevatron has completed its short-term search for dilepton and trilepton
production.
\end{abstract}

\vspace{0.5cm}
\begin{flushleft}
\baselineskip=12pt
{CERN-TH.7535/94}\\
{CTP-TAMU-66/94}\\
{ACT-23/94}\\
December 1994
\end{flushleft}
\newpage

\setcounter{page}{1}
\pagestyle{plain}
\baselineskip=14pt

Experimental searches for supersymmetric particles have come a long way since
the commissioning of the Tevatron $p\bar p$ collider at Fermilab (1988) and the
LEP $e^+e^-$ collider at CERN (1989). The strengths and weaknesses of these
two types of colliders are well known. A hadron collider is best suited for
searching the highest accessible mass scales since a sharp kinematical limit
does not exist, but discoverability depends on the event rate and the
cleanliness of the signal. An $e^+e^-$ collider is capable of discovery
essentially up to the kinematical limit, but this is much lower than what would
be accessible in a hadron collider. In fact, it has become apparent that an
$e^+e^-$ linear collider with a center-of-mass energy in the multi-hundred
GeV range would be ideal for what has been termed ``sparticle spectroscopy".
At present though, in the search for new physics we have to make the best
possible use of existing facilities, since information gathered there would
illuminate the path towards higher energy machines. One such effort is being
conducted at the Tevatron, where the search for weakly interacting sparticles
(charginos and neutralinos) has become quite topical, in view of the fact that
the reach of the machine for the traditional strongly interacting sparticles
has been nearly reached. This effort will benefit from an integrated luminosity
in excess of $100\ipb$ by the end of the ongoing Run IB, and possibly
1--2$\ifb$ during the Main Injector era around the year 2000.

In this paper we study the prospects for supersymmetry discovery at the
Tevatron via the hadronically quiet dilepton and trilepton
signals\footnote{These signals contain no hadronic activity, except for
initial state radiation effects, and are thus distinct from the usual
multilepton signals in squark and gluino production.} which occur in the
production and decay of charginos and neutralinos in $p\bar p$ collisions
\cite{trileptons,BT,LNWZ}. Our previous study \cite{LNWZ} considered the
trilepton signal with an estimated $100\ipb$ of accumulated data. Here we
update this analysis by incorporating the latest experimental information on
the trilepton signal and prospects for its detection with  ${\cal O}(1\ifb)$
data. We also include for the first time the dilepton signal, which has been
recently shown to be experimentally extractable \cite{James}, and as we
discuss, has the advantage of allowing a significant exploration of the
parameter space for chargino masses in the LEPII accessible range. Our
calculations are performed in the context of a broad class of unified
supergravity models, which have the virtue of having the least number of
free parameters and are therefore highly predictive and straightforwardly
testable through a variety of correlated phenomena at different experimental
facilities. Our study should give a good idea of the range of possibilities
open to experimental investigation, and allow quantitative checks of specific
models which yield the largest rates.

We consider unified supergravity models with universal soft supersymmetry
breaking at the unification scale, and radiative electroweak symmetry breaking
(enforced using the one-loop effective potential) at the weak scale
\cite{reviews}. These constraints reduce the number of parameters needed to
describe the models to four, which can be taken to be $m_{\chi^\pm_1},
\xi_0\equiv m_0/m_{1/2},\xi_A\equiv A/m_{1/2},\tan\beta$, with a specified
value for the top-quark mass ($m_t$). In what follows we take $m_t^{\rm
pole}=160\GeV$ which is the central value obtained in fits to all electroweak
and Tevatron data in the context of supersymmetric models~\cite{EFL}. Of
relevance to our discussion, we note that in all models considered
the following relation holds to various degrees of approximation
\begin{equation}
m_{\chi^\pm_1}\approx m_{\chi^0_2}\approx2m_{\chi^0_1}\ .
\label{masses}
\end{equation}
Among these four-parameter supersymmetric models we consider generic models
with continuous values of $m_{\chi^\pm_1}$ and discrete choices for the other
three parameters:
\begin{equation}
\tan\beta=2,10\ ;\qquad \xi_0=0,1,2,5\ ;\qquad \xi_A=0\ .
\label{generic}
\end{equation}
The choices of $\tan\beta$ are representative; higher values of $\tan\beta$ are
likely to yield values of $B(b\to s\gamma)$ in conflict with present
experimental limits \cite{LargeTanB}. The choices of $\xi_0$ correspond to
$m_{\tilde q}\approx(0.8,0.9,1.1,1.9)m_{\tilde g}$. The choice of $A$ has
little impact on the results. We also consider the case of minimal $SU(5)$
supergravity, where in addition we impose the constraints from proton decay
and cosmology (a not too young Universe). The parameter space in this case
is still four-dimensional, but restricted to $\tan\beta\lsim10$, $\xi_0\gsim4$,
and $m_{\chi^\pm_1}\lsim120\GeV$ \cite{LNPetc}.

We also consider the case of no-scale $SU(5)\times U(1)$ supergravity
\cite{reviews}. In this class of models the supersymmetry breaking parameters
are related in a string-inspired way. In the two-parameter {\em moduli}
scenario $\xi_0=\xi_A=0$ \cite{LNZI}, whereas in the {\em dilaton} scenario
$\xi_0={1\over\sqrt{3}},\ \xi_A=-1$ \cite{LNZII}. We also compute the rates in
the {\em one-parameter} moduli ($B(M_U)=0$) and dilaton ($B(M_U)=2m_0$)
scenarios, where $M_U$ is the string unification scale, and with this extra
condition $\tan\beta$ is determined as a function of $m_{\chi^\pm_1}$, which
is the only free parameter in the model. A series of experimental constraints
and predictions for these models have been given in Refs.~\cite{Easpects} and
\cite{One}, respectively.
\vspace{0.5cm}

The processes of interest are
\begin{itemize}
\item {\em Trileptons}: $p\bar p\to \chi^0_2\chi^\pm_1$, where the
next-to-lightest neutralino decays leptonically
($\chi^0_2\to\chi^0_1\ell^+\ell^-$), and so does the lightest chargino
($\chi^\pm_1\to\chi^0_1\ell^\pm\nu_l$). The cross section proceeds
via $s$-channel exchange of an off-shell $W$ and (small) $t$-channel squark
exchange, and thus peaks at $m_{\chi^\pm_1}\approx {1\over2}M_W$, and otherwise
falls off smoothly with increasing chargino masses with a small $\tan\beta$
dependence.
\item {\em Dileptons}: $p\bar p\to\chi^+_1\chi^-_1$, where both charginos decay
leptonically. The cross section proceeds via $s$-channel exchange of off-shell
$Z$ and $\gamma$ and $t$-channel squark exchange, and peaks for
$m_{\chi^\pm_1}\approx{1\over2}M_Z$. Dileptons could also come from
$p\bar p\to\chi^0_1\chi^0_2,\chi^0_2\chi^0_2$, with the appropriate leptonic
or invisible decays of $\chi^0_2$. Both of these processes are negligible
\cite{BT} because the couplings of the $Z$ and $\gamma$ to neutralinos are
highly suppressed when the neutralinos have a high gaugino content, as is the
case when Eq.~(\ref{masses}) holds. Yet another source of dileptons via $p\bar
p\to \tilde e^+_R\tilde e^-_R$ suffers from small rates for selectron masses
beyond the LEP limit~\cite{BCPT}.
\end{itemize}

The more important factors in the dilepton and trilepton yields are the
leptonic branching fractions which can vary widely throughout the parameter
space \cite{LNWZ}. If all sparticles are fairly heavy, the decay amplitude is
dominated by $W$ or $Z$ exchange. In this case the branching fractions into
electrons plus muons are $B(\chi^\pm_1\to\chi^0_1\ell^\pm\nu_l)\approx2/9$ and
$B(\chi^0_2\to\chi^0_1\ell^+\ell^-)\approx6\%$. On the other hand, if some of
the sparticles are relatively light, most likely the sleptons, the branching
fractions are altered. The extreme, although not unusual, case occurs when
the sleptons are on-shell. These two-body decays then dominate and the chargino
leptonic branching fraction is maximized, \ie,
$B(\chi^\pm_1\to\chi^0_1\ell^\pm\nu_l)_{\rm max}=2/3$. Light sleptons also
affect the neutralino leptonic branching ratio.\footnote{In supergravity
models $m_{\tilde e_R}=m_{\tilde\mu_R}<m_{\tilde e_L}=m_{\tilde\mu_L}$.}
When the sneutrino is on-shell
and is lighter than the corresponding right-handed charged slepton ($\tilde
e_R,\tilde\mu_R$), the channel $\chi^0_2\to \nu_l\tilde\nu_l$  dominates the
amplitude, and the neutralino leptonic branching ratio is suppressed. This
situation is reversed when the charged slepton is on-shell and is lighter than
the sneutrino, which leads to an enhancement of the neutralino leptonic
branching ratio. For sufficiently high neutralino and chargino masses, both
leptonic branching ratios decrease because the $W$ and $Z$ go on-shell and
dominate the decay amplitudes. In the case of the neutralino, the spoiler mode
$\chi^0_2\to \chi^0_1 h$ also becomes kinematically allowed. These high-mass
suppressions do not kick in until chargino and neutralino masses
$m_{\chi^\pm_1}\approx m_{\chi^0_2}\sim 2M_Z,2m_h\sim200\GeV$.

The results of our computations for the various models are shown in
Figs.~\ref{di-trilep.SSM2.eps},\ref{di-trilep.SSM10.eps},
\ref{di-trilep.min.eps},\ref{di-trilep.nsc.eps},\ref{di-trilep.kl.eps},\ref{di-trilep.B0Bnew.eps}. The various curves in the figures terminate at the low end
because of various parameter space constraints, whereas at the high end they
are cutoff when the yields fall below the foreseeable sensitivity.\footnote{For
$\xi_0=5$, radiative electroweak symmetry breaking is only possible for
$\tan\beta\lsim4$. This is why there is no curve for $\xi_0=5$ in
Fig.~\ref{di-trilep.SSM10.eps} ($\tan\beta=10$), whereas there is such a curve
in Fig.~\ref{di-trilep.SSM2.eps} ($\tan\beta=2$).} In most cases we note that
the rates are higher for $\mu<0$. This is a consequence of suppressed branching
fractions for $\mu>0$, but also a generally smaller allowed parameter space
which requires minimum values of the chargino mass which may exceed
significantly the present experimental lower limit. We can also observe that
the dilepton rates indeed peak near ${1\over2}M_Z$, whereas the trilepton rates
are not as large for light chargino masses since they peak at ${1\over2}M_W$.
It is also evident in the figures that for chargino masses below $\sim100\GeV$,
the trilepton rates can be highly suppressed, while the dilepton rates are not,
thus producing a rather complementary effect. This ``threshold" phenomenon is
most evident in
Figs.~\ref{di-trilep.nsc.eps},\ref{di-trilep.kl.eps},\ref{di-trilep.B0Bnew.eps}
and, as dicussed above, corresponds to a suppression of the neutralino leptonic
branching ratio for light sleptons, \ie, when $\chi^0_2\to\nu\tilde\nu$ is
allowed.

\begin{table}[t]
\begin{center}
\caption{Estimated chargino mass reaches in various supergravity models for
chargino-neutralino production in $p\bar p$ collisions at the Tevatron via
dilepton and trilepton modes for integrated luminosities of $100\ipb$ and
$1\ifb$. All masses in GeV. Dashes (--) indicate negligible sensitivity.}
\label{Table}
\smallskip
\begin{tabular}{|c|c|c|c|c|c|}\hline
\multicolumn{2}{|c|}{generic}&\multicolumn{2}{c|}{$\mu>0$}
&\multicolumn{2}{c|}{$\mu<0$}\\ \hline
$\tan\beta$&$\xi_0$&$100\ipb$&$1\ifb$&$100\ipb$&$1\ifb$\\ \hline
2&0&--&120&100&145\\
&1&--&125&75&115\\
&2&--&100&65&100\\
&5&--&80&55&80\\ \hline
10&0&--&105&70&135\\
&1&70&95&65&100\\
&2&--&70&--&70\\ \hline
\end{tabular}
\medskip

\begin{tabular}{|c|c|c|c|c|c|}\hline
\multicolumn{2}{|c|}{Model}&\multicolumn{2}{c|}{$\mu>0$}
&\multicolumn{2}{c|}{$\mu<0$}\\ \hline
&$\tan\beta$&$100\ipb$&$1\ifb$&$100\ipb$&$1\ifb$\\ \hline
moduli&2&--&115&100&150\\
(2-par)&6&75&160&100&150\\
&10&75&160&70&150\\ \hline
dilaton&2&95&135 &80&120\\
(2-par)&6&80&130&80&120\\
& 10&80&125&80&120\\ \hline
\multicolumn{2}{|c|}{moduli (1-par)}&N/A&N/A&70&150\\ \hline
\multicolumn{2}{|c|}{dilaton (1-par)}&N/A&N/A&80&125\\ \hline
\multicolumn{2}{|c|}{minimal $SU(5)$}&50&80&50&75\\ \hline
\end{tabular}
\end{center}
\hrule
\end{table}

The significance of our results is quantified by the horizontal dashed lines in
the figures, which represent estimates of the experimental sensitivity to be
reached with $100\ipb$ (upper lines) and $1\ifb$ (lower lines). The lesser
sensitivity should be achievable at the end of Run IB during 1995 (\ie, prior
to the LEPII upgrade), whereas the higher sensitivity should be available
with the Main Injector upgrade in 1999 (\ie, after the LEPII shutdown but
before the LHC commissioning).

The trilepton sensitivity with $100\ipb$ (\ie, $0.4\pb$) has been estimated by
simply scaling down by a factor of 5 the present experimental limit of
$\sim2\pb$ obtained with $20\ipb$ of recorded data \cite{Kato}. The factor of 5
is the expected increase in recorded luminosity, and a simple ${\cal L}$
scaling is appropriate assuming the trilepton signal has no Standard Model
backgrounds at this level of sensitivity. The sensitivity at $1\ifb$ requires a
study of the background since small Standard Model processes and
detector-dependent instrumental backgrounds become important at this level of
sensitivity \cite{DiTevatron}. The sensitivity in the figures (\ie, $0.07\pb$)
is obtained by scaling up by $\sqrt{\cal L}$ the value given in Table~II of
Ref.~\cite{DiTevatron}.

The dilepton (plus $\mpT$) signal suffers from several Standard Model
backgrounds, most notably $Z\to\tau\tau$ and $WW$ production. A study based on
the D0 detector \cite{James} reveals that with suitable cuts, in $100\ipb$
an estimated background of 8 events is expected, which would require 8 signal
events at $3\sigma$ significance. The efficiencies for dilepton detection have
also been studied \cite{James}, and they improve with increasing chargino
masses, 8\% is a typical value. All this implies a sensitivity of $1\pb$ for
dilepton detection. With $1\ifb$ one can scale down the sensitivity with
$\sqrt{\cal L}$, obtaining a sensitivity of $0.3\pb$.

The reaches in chargino masses in the various models, can be readily obtained
from the figures by considering both dilepton and trilepton signals, and are
summarized in Table~\ref{Table} for the two integrated luminosity scenarios.
The reaches in Table~\ref{Table} translate into indirect reaches
in every other sparticle mass, since they are all related. In particular,
$m_{\chi^\pm_1}\sim 0.3m_{\tilde g}$ and
$m_{\tilde q}\approx (m_{\tilde g}/2.9)\sqrt{6+\xi^2_0}$ (in the $SU(5)\times
U(1)$ models the numerical coefficients in this relation are slightly
different, implying $m_{\tilde q}\approx m_{\tilde g}$). It is also interesting
to point out that the pattern of yields for the various models is quite
different, therefore observation of a signal will disprove many of the models,
while supporting a small subset of them.

It has been pointed out that the dilepton and trilepton data sample may be
enhanced by considering presumed trilepton events where one of the leptons
is either missed or has a $p_T$ below 5 GeV (``2-out-of-3") \cite{James2}.
Such enhancements would alter our reach estimates above, making them even more
promising.

{}From Table~\ref{Table} it is clear that in some regions of parameter space,
the reach of the Tevatron for chargino masses is quite significant. With
$100\ipb$ it should be possible to probe chargino masses as high as 100 GeV
in the generic models for $\tan\beta=2$, $\xi_0=0$, and $\mu<0$, and in the
two-parameter $SU(5)\times U(1)$ moduli scenario for $\tan\beta\lsim10$. More
generally, the accessible region of parameter space should overlap with that
within the reach of LEPII, although it would have been explored before LEPII
turns on. However, LEPII has an important task: chargino searches at LEPII will
not be hindered by small branching fractions, and thus a more model-independent
lower limit on the chargino mass should be achievable, \ie,
$m_{\chi^\pm_1}\lsim{1\over2}\sqrt{s}$.\footnote{At LEPII it might be possible
to extend the indirect reach for charginos by studying the process
$e^+e^-\to\chi^0_1\chi^0_2$ with $\chi^0_2\to\chi^0_1+2j$.
Equation~(\ref{masses}) implies a kinematical reach of
$m_{\chi^\pm_1}\lsim{2\over3}\sqrt{s}$.}  We would like to
conclude with Fig.~\ref{di-trilep.B0Bnew.eps}, where we show the predictions
for the dilepton and trilepton rates in our chosen one-parameter $SU(5)\times
U(1)$ models, which are the most predictive supersymmetric models to date.
It is interesting to note that in the moduli scenario, the mass reach for
charginos could be as high as $150\GeV$ with an integrated luminosity of
$1\ifb$. Further proposed increases in luminosity or center-of-mass energy of
the Tevatron collider have the potential of probing even deeper into the
parameter space \cite{DiTevatron}. We conclude that detection of weakly
interacting sparticles at the Tevatron may well bring the first direct signal
for supersymmetry.

\section*{Acknowledgments}
We would like to thank James White for motivating this study and for providing
us with valuable information about the sensitivity of the dilepton signal.
This work has been supported in part by DOE grant DE-FG05-91-ER-40633. The
work of X. W. has been supported by the World Laboratory.

\newpage

\begin{figure}[p]
\vspace{6in}
\includegraphics{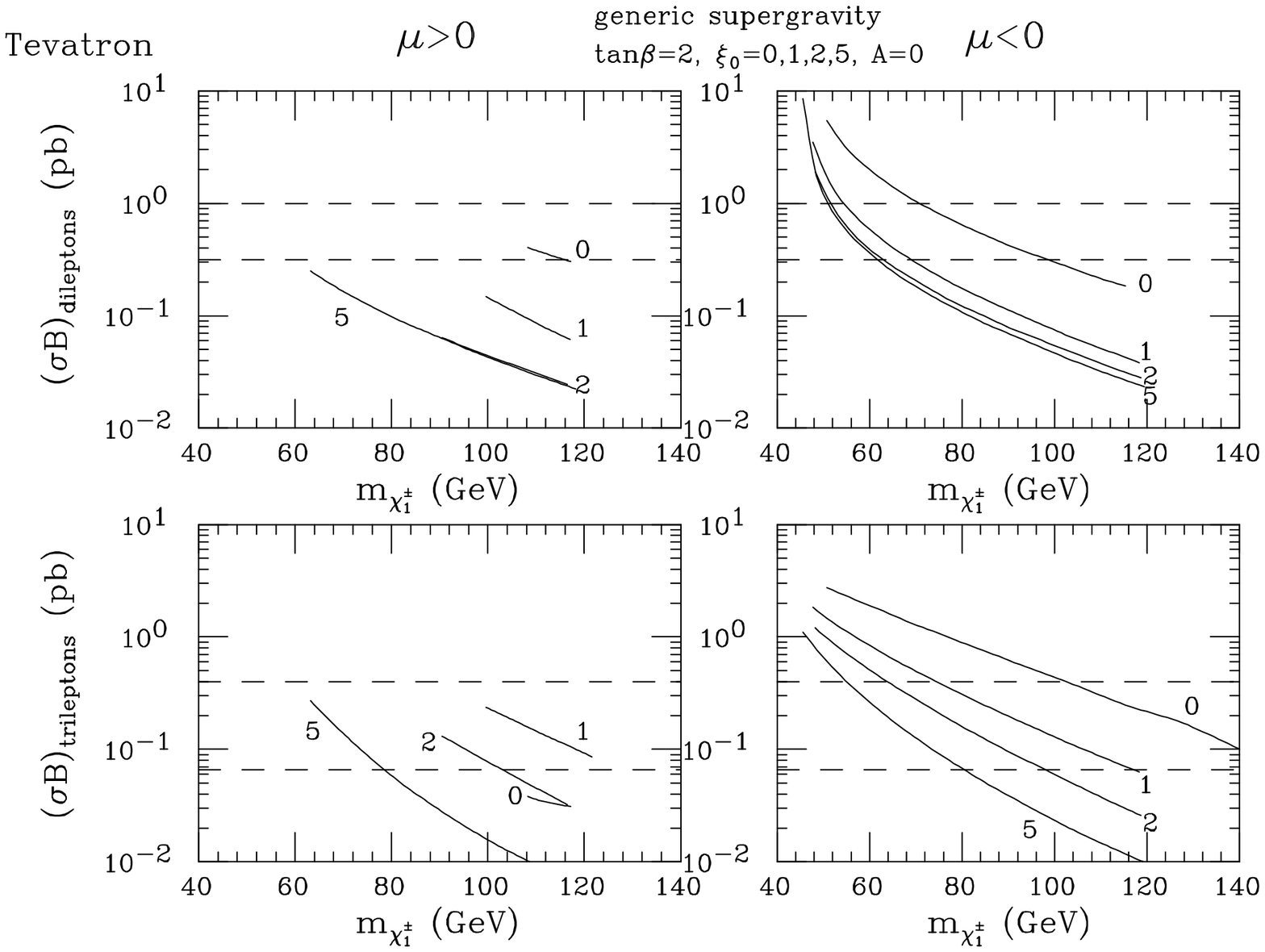}
\caption{The dilepton and trilepton rates at the Tevatron versus the chargino
mass in a generic unified supergravity model with $\tan\beta=2$,
$\xi_0=0,1,2,5$ (as indicated), and $A=0$. The upper (lower) dashed lines
represent estimated reaches with $100\,{\rm pb}^{-1}$ ($1\,{\rm fb}^{-1}$) of
data.}
\label{di-trilep.SSM2.eps}
\end{figure}
\clearpage

\begin{figure}[p]
\vspace{6in}
\includegraphics{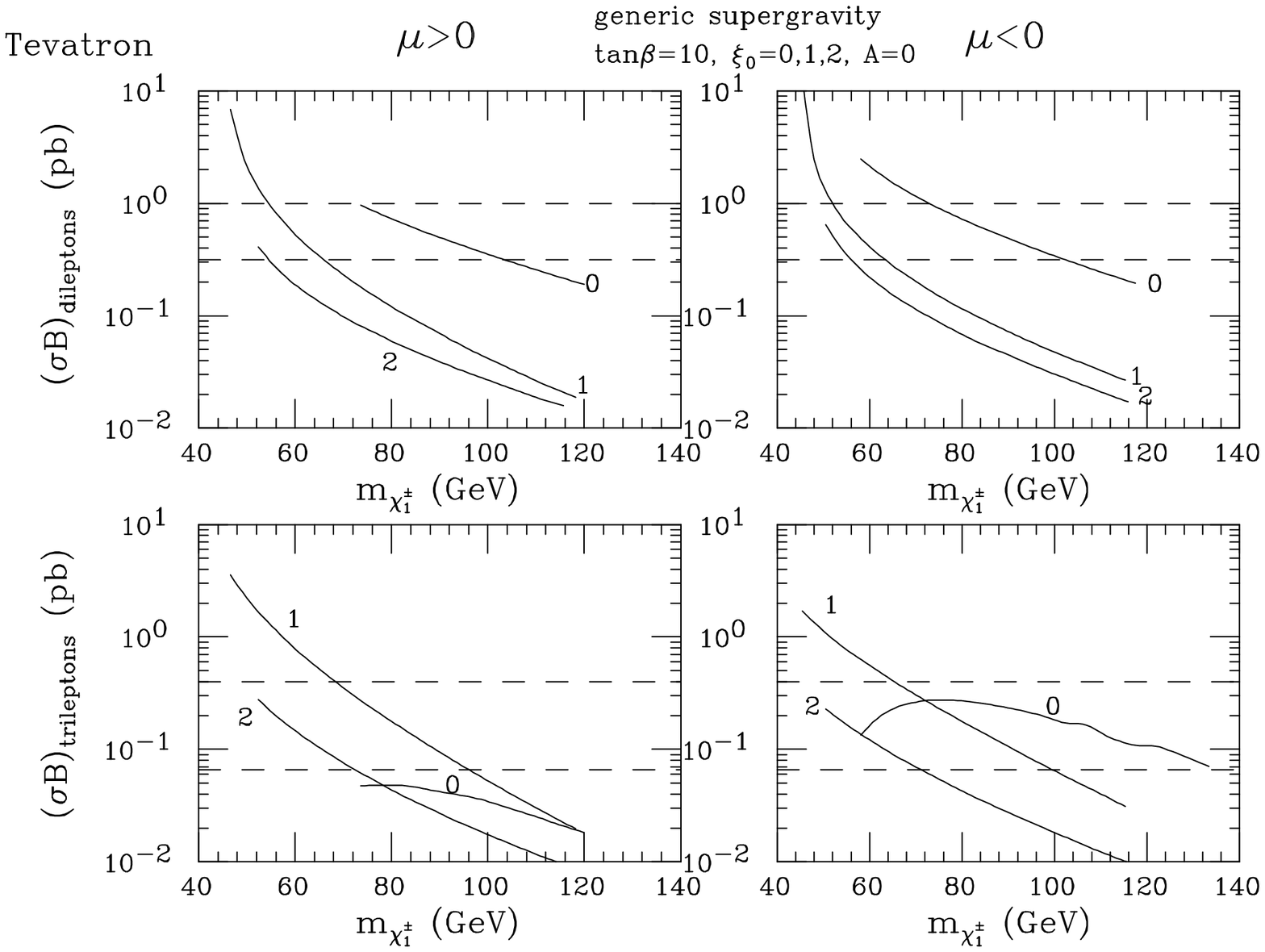}
\caption{The dilepton and trilepton rates at the Tevatron versus the chargino
mass in a generic unified supergravity model with $\tan\beta=10$, $\xi_0=0,1,2$
(as indicated), and $A=0$. The upper (lower) dashed lines represent estimated
reaches with $100\,{\rm pb}^{-1}$ ($1\,{\rm fb}^{-1}$) of data.}
\label{di-trilep.SSM10.eps}
\end{figure}
\clearpage

\begin{figure}[p]
\vspace{6in}
\includegraphics{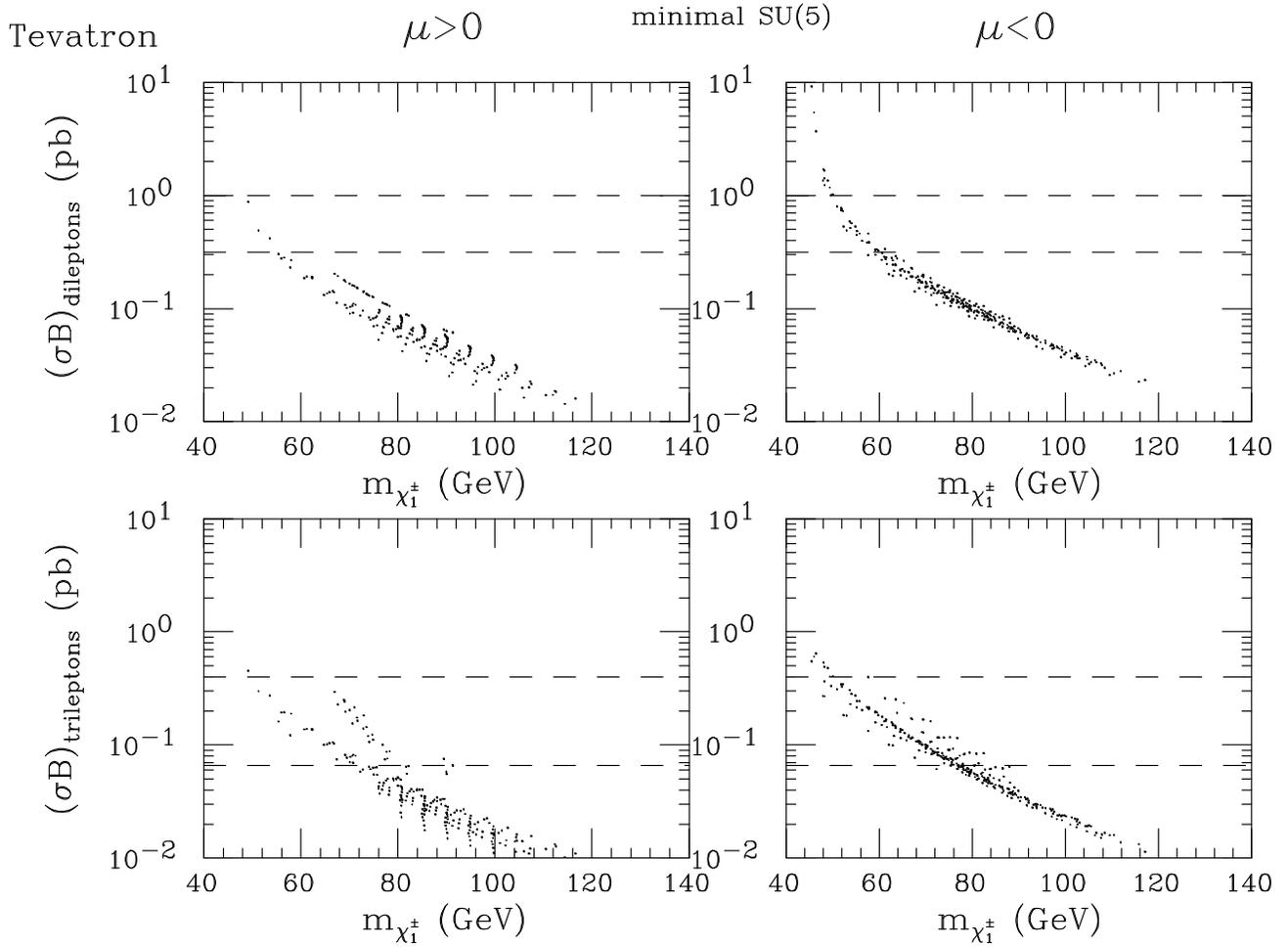}
\caption{The dilepton and trilepton rates at the Tevatron versus the chargino
mass in the minimal $SU(5)$ supergravity model (where $\tan\beta<10$,
$\xi_0>4$). The upper (lower) dashed lines represent estimated reaches with
$100\,{\rm pb}^{-1}$ ($1\,{\rm fb}^{-1}$) of data.}
\label{di-trilep.min.eps}
\end{figure}
\clearpage

\begin{figure}[p]
\vspace{6in}
\includegraphics{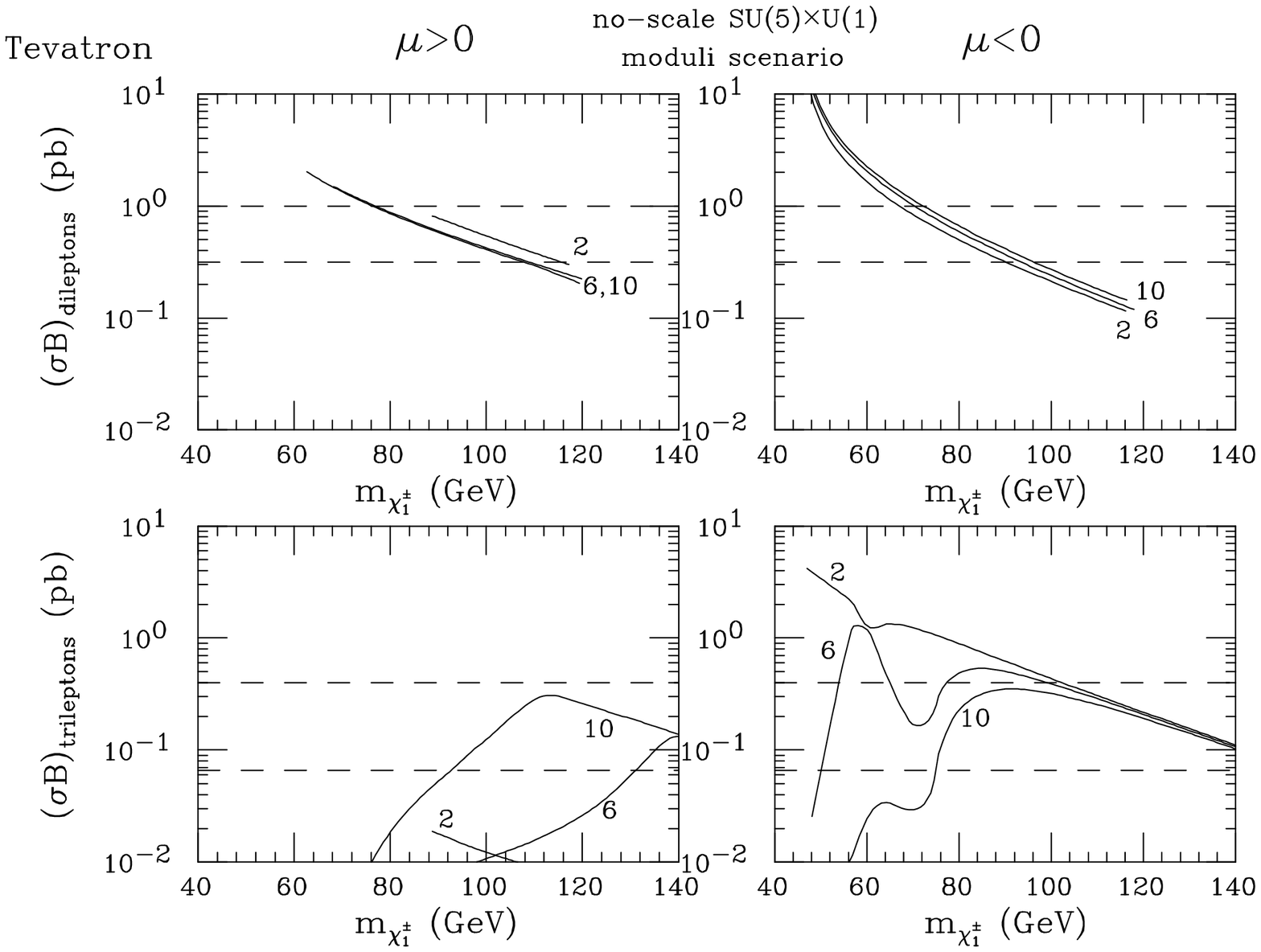}
\caption{The dilepton and trilepton rates at the Tevatron versus the chargino
mass in two-parameter $SU(5)\times U(1)$ supergravity -- moduli scenario
($\xi_0=\xi_A=0$) for the indicated values of $\tan\beta$. The upper (lower)
dashed lines represent estimated reaches with $100\,{\rm pb}^{-1}$ ($1\,{\rm
fb}^{-1}$) of data.}
\label{di-trilep.nsc.eps}
\end{figure}
\clearpage

\begin{figure}[p]
\vspace{6in}
\includegraphics{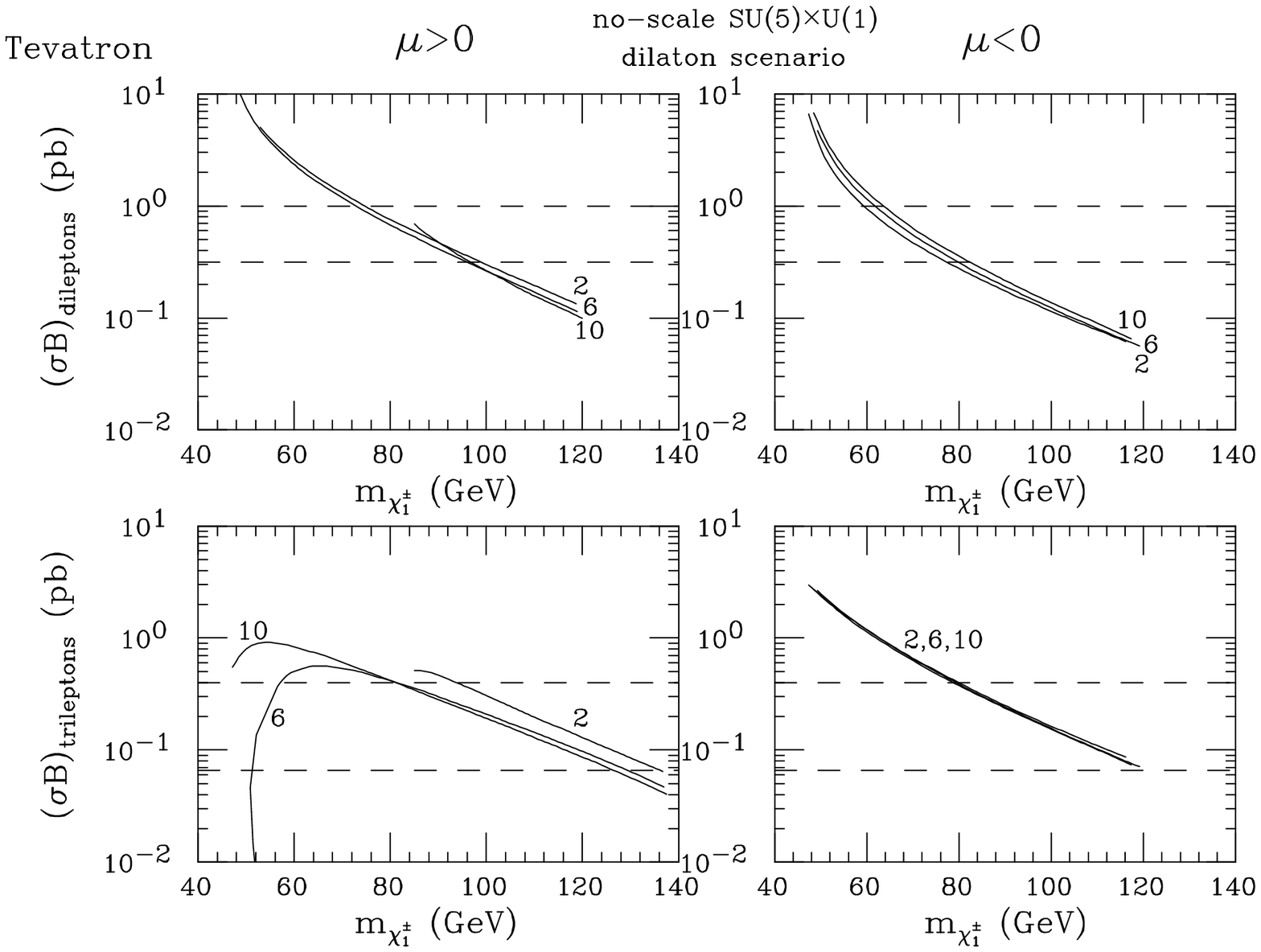}
\caption{The dilepton and trilepton rates at the Tevatron versus the chargino
mass in two-parameter $SU(5)\times U(1)$ supergravity -- dilaton scenario
($\xi_0={1\over\surd3},\xi_A=-1$) for the indicated values of $\tan\beta$. The
upper (lower) dashed lines represent estimated reaches with $100\,{\rm
pb}^{-1}$ ($1\,{\rm fb}^{-1}$) of data.}
\label{di-trilep.kl.eps}
\end{figure}
\clearpage

\begin{figure}[p]
\vspace{6in}
\includegraphics{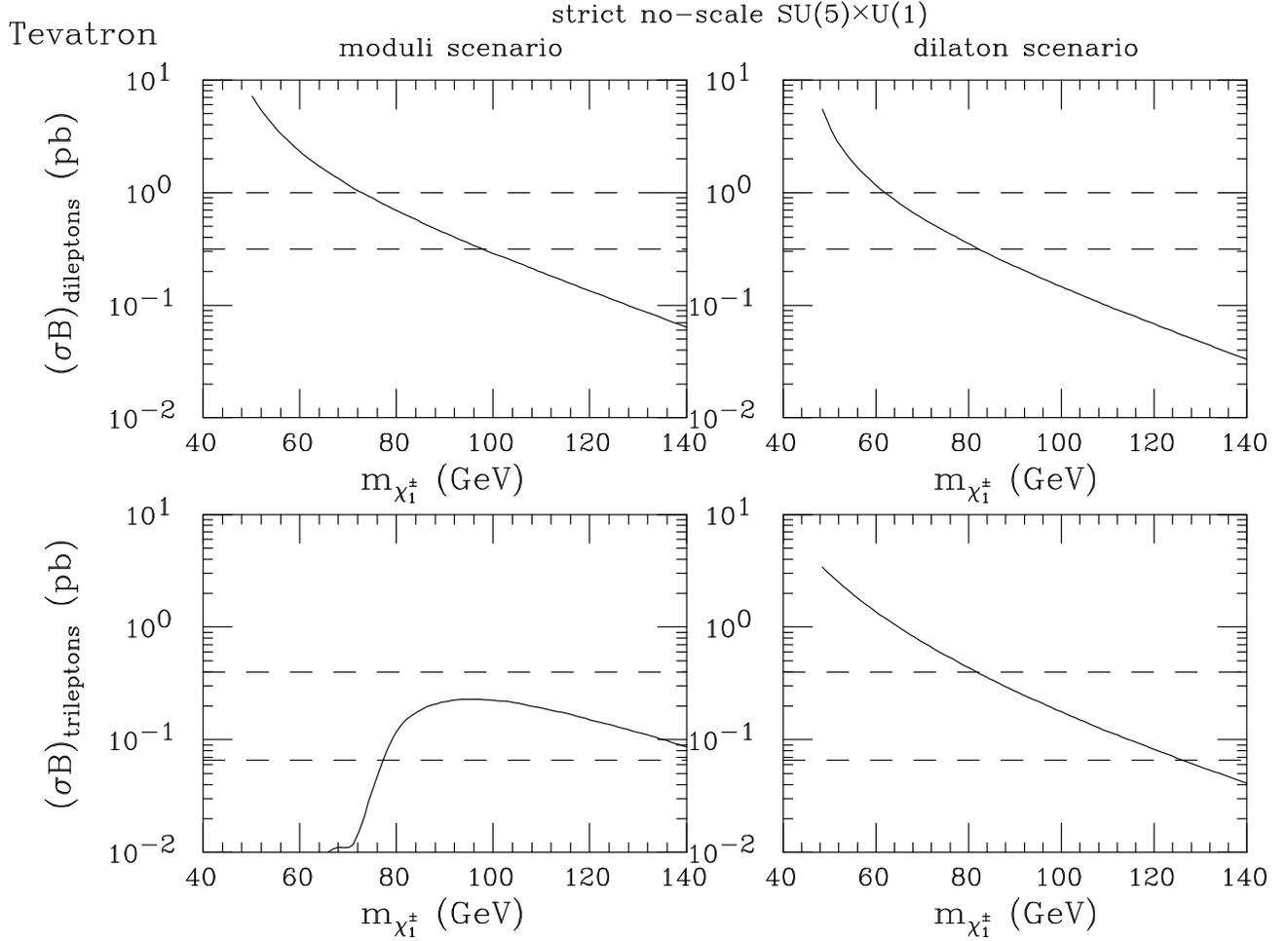}
\caption{The dilepton and trilepton rates at the Tevatron versus the chargino
mass in one-parameter $SU(5)\times U(1)$ supergravity -- moduli and dilaton
scenarios ($\mu<0$ in both cases). The upper (lower) dashed lines represent
estimated reaches with $100\,{\rm pb}^{-1}$ ($1\,{\rm fb}^{-1}$) of data.}
\label{di-trilep.B0Bnew.eps}
\end{figure}
\clearpage

\end{document}